\documentclass[lettersize,journal]{IEEEtran}

\usepackage{hyperref}
\usepackage{latexsym}
\usepackage{graphicx}
\usepackage{epsfig}
\usepackage{url}
\usepackage{amsmath,amssymb}
\usepackage{latexsym}
\usepackage{xspace}
\usepackage{authblk}
\usepackage{algorithm}
\usepackage{algpseudocode}
\usepackage{algorithmicx}

\usepackage{listings}
\algnewcommand\algorithmicforeach{\textbf{for each}}
\algdef{S}[FOR]{ForEach}[1]{\algorithmicforeach\ #1\ \algorithmicdo}

\usepackage{soul}
\usepackage{multirow}
\usepackage{color, soul}
\usepackage{comment}
\usepackage{enumitem}
\usepackage{array,tabularx}
\setlist[enumerate]{%
wide =0.6\parindent}%

\usepackage[colorinlistoftodos]{todonotes}

\usepackage{framed,enumitem}

\begin{document}
\title{The SemIoE Ontology: A Semantic Model Solution for an IoE-based Industry}

\author{Marco Arazzi$^{1}$, Antonino Nocera$^{1}$, Emanuele Storti$^{2}$\\
$^{1}$ Department of Electrical, Computer and Biomedical Engineering, University of Pavia \\
via A. Ferrata, 5 - 27100 - Pavia, Italy.
\\
marco.arazzi01@universitadipavia.it; antonino.nocera@unipv.it \\
$^{2}$ Department of Information Engineering, Polytechnic University of Marche \\
via Brecce Bianche - 60121 - Ancona, Italy\\
e.storti@univpm.it}

\date{}

\maketitle

\begin{abstract}
Recently, the Industry 5.0 is gaining attention as a novel paradigm, defining the next concrete steps toward more and more intelligent, green-aware and user-centric digital systems.

In an era in which smart devices typically adopted in the industry domain are more and more sophisticated and autonomous, the Internet of Things and its evolution, known as the Internet of Everything (IoE, for short), involving also people, robots, processes and data in the network, represent the main driver to allow industries to put the experiences and needs of human beings at the center of their ecosystems.
However, due to the extreme heterogeneity of the involved entities, their intrinsic need and capability to cooperate, and the aim to adapt to a dynamic user-centric context, special attention is required for the integration and processing of the data produced by such an IoE.
This is the objective of the present paper, in which we propose a novel semantic model that formalizes the fundamental actors, elements and information of an IoE, along with their relationships.
In our design, we focus on state-of-the-art design principles, in particular reuse, and abstraction, to build ``SemIoE'', a lightweight ontology inheriting and extending concepts from well-known and consolidated reference ontologies.

The defined semantic layer represents a core data model that can be extended to embrace any modern industrial scenario. It represents the base of an IoE Knowledge Graph, on top of which, as an additional contribution, we analyze and define some essential services for an IoE-based industry.
\end{abstract}

\section{Introduction}
\IEEEPARstart{T}{he} Industry 5.0 paradigm emphasizes the shift from traditional industries to intelligent, people-oriented, and environmentally conscious industrial ecosystems.
According to this perspective, the focus of production processes should always revolve around the experiences and needs of human beings, fostering innovation and creativity on the production side, and individualizing product offerings on the customer side. 
To achieve these goals, production environments must undergo significant adjustments to delegate repetitive tasks to more and more autonomous devices and machines. The Internet of Things has strengthened the relationship between humans and intelligent devices, and the expansion of this network, known as the Internet of Everything (IoE), integrates devices, processes, data, and people into a single, highly interconnected environment.
This technology allows for a smooth transition towards Industry 5.0, as the IoE natively builds a collaborative ecosystem involving people, autonomous devices (such as smart objects, robots, cobot, and so forth), and smart platforms.
However, one of the main challenges of an Industry 5.0 scenario is to re-define the working context by putting human resources at the center of the production chain, so that all the services, facilities, data, and tools, available in the industry environment, are dynamically adapted to his/her needs. 
Under this point of view, if the IoE becomes a technological opportunity to build an Industry 5.0 on the one hand, suitably modeling its data and building human-centric services on top of them is a critical task, on the other hand.
Moreover, modern IoT technology, which is part of the more general IoE ecosystem considered in this paper, provides access to increasingly autonomous smart devices that can often even act on behalf of their human owners. This leads to a very complex reality to which the industry should adapt, not only to surround human resources easing their access to the industrial environment, but also to allow a smooth shift from direct manual actions and tasks carried out by human resources to automatic actions carried out by suitably delegated smart devices.
Such an opportunity implicitly requires a slight paradigm change from human-centric to user-centric, in which users can be either human resources or autonomous smart devices acting on their behalves.

In such a scenario one of the main problems is how to integrate data produced by intelligent and autonomous devices with those produced by the actions carried out by humans in order to enable self-organization, self-optimization, self-healing for the whole IoE. 
To address the above mentioned problem, large integration scenarios must be considered, up to the level of an entire enterprise, for collecting and processing flows of data produced by all the agents of an industrial IoE. 
In this setting, a model-based strategy may support the integration effort providing fundamental reference information to enable the declination of the above mentioned Industry 5.0, at least from a data management point of view.
However, such a strategy should answer the following research question: how to model heterogeneous information from such a complex context in a flexible, modular and extensible way?

In this paper, we embrace the novel scenario introduced above and strive to address the  mentioned 
question by introducing a user-centric {\em semantic} model that formalizes and relates the fundamental actors, elements and information available in such an IoE environment for an Industry 5.0 ecosystem. After a thorough analysis of the reference domain, we recognize the following main entities for our model: {\em (i)} human resources and smart devices, globally referred to as {\em agents} of the IoE in the remaining of this paper, {\em (ii)} activities and workflows carried out by IoE agents, {\em (iii)} relationships and collaboration actions among agents, {\em (iv)} role and activity-related privileges, {\em (v)} environment elements and physical building parts. On top of these concepts, a lightweight and extensible ontology is designed following state-of-the-art design principles, in particular reuse, and abstraction.
As for this last aspect, the ontology is built by inheriting concepts available in well-supported and acknowledged existing models, such as the Semantic Sensor Network (SOSA/SSN) \cite{compton2012ssn}, the Building Topology Ontology (BOT) \cite{rasmussen2021bot}, and the Organization Ontology (ORG) \cite{world2014organization}.
The proposed lightweight ontology, named ``SemIoE'', represents a semantic knowledge core layer providing a reference structure to build an IoE Knowledge Graph, which includes all the instances of the entities identified above for an Industry 5.0.
Due to the extremely dynamic and heterogeneous domain, instead of focusing on specific industry ecosystems, we make an additional effort to generalize and identify the common actors and build a common core model, which can, hence, be extended to model any existing real-life industrial IoE scenario.
As an additional important contribution, in our solution we define and include a set of services for the IoE built on top of the aforementioned IoE knowledge Graph, namely: {\em (i)} IoE Access Control, managing and securing the access to IoE facilities only to sufficiently privileged agents, {\em (ii)} IoE Collaboration, providing a robust mechanism to enable collaboration among IoE agents towards the completion of specific IoE tasks, {\em (iii)} IoE Secure Delegation, enabling the possibility of delegating complex tasks to supporter peers (both humans and smart devices), {\em (iv)} IoE Environment Setting, allowing for the adaptation of the surrounding smart environment to favor the activities/needs of IoE agents.
Altogether, the semantic knowledge layer, the IoE Knowledge Graph, and the set of defined IoE services represent the building blocks for an IoE-based Industry 5.0.

The remaining of this paper is organized as follows.
In Section \ref{sec:RW}, we analyze the related scientific literature. The reference scenario is detailed in Section \ref{sec:scenario}. Section \ref{sec:Ontology} is devoted to the description of the proposed semantic knowledge layer and the discussion of the proposed ontological model for IoE. In Section \ref{sec:services}, we describe in details the services for the IoE included in our solution. Section \ref{sec:eval}, instead, is devoted to analyze possible implementation strategies. Finally, in Section \ref{sec:conclusion} we draw our conclusion and discuss possible open points and future developments.

\section{Related Work}
\label{sec:RW}
\noindent
The novel paradigm of Industry 5.0 identifies the transition from the traditional industry toward smart, eco-aware, and human-centric factories \cite{maddikunta2022industry,xu2021industry}.
In this futuristic, but still timely, context, some authors identify the Internet of Things and, more in general, the Internet of Everything, as an enabling technology \cite{leng2022industry}.
The Internet of Everything focuses on the interconnection between humans, data processes and things in a single unified ecosystem and is referred to as the evolution of the consolidated concept of the Internet of Things (IoT) \cite{rm2020load}.
Due to the heterogeneity of the technologies used in IoT, and the additional complexity of combining data produced by IoT devices with those produced by humans, processes, and complex interaction among them, it is fundamental to define ontologies allowing semantic interoperability between different applications and services \cite{rahman2020comprehensive}.

In the literature, a great effort has been devoted to defining  semantic models to represent device characteristics and their relationships in the Internet of Things.
A major problem faced by the research community in this context is related to the heterogeneity of IoT devices and their continuous evolution over time.
As a matter of fact, ontologies in this domain should be general enough to model network of sensors, sensors' capabilities and the applications built on top of them. Moreover, they should also consider that these technologies are extremely mutable, as new devices are released on a daily basis, and that the difference between the applications and services in different industrial domain are, typically, very relevant \cite{schlenoff2013literature}.
One of the earliest approaches in this direction is described in \cite{russomanno2005building}. Here, the authors designed a very general-purpose ontology, called OntoSensor, starting from the Web Ontology Language \cite{mcguinness2004owl} and the Suggested Upper Merged Ontology (SUMO) \cite{pease2002suggested}.
Building once again from the SUMO ontology, the authors of \cite{eid2006novel} designed an initial ontology to retrieve all-and-only relevant sensor data, following an evolving prototype life-cycle. According to the approach identified in \cite{horridge2004practical}, the building process is split into: common vocabulary collection, initial taxonomy identification, adding restrictions and axioms, checking for consistency, modifications, and evaluation.

At the time of writing, probably the most suitable sensor ontology for IoT is the Semantic Sensor Network (SSN) ontology \cite{compton2012ssn}.
Several work built lightweight semantic models on top of the concepts included in the SSN ontology, e.g. \cite{bermudez2016iot,JANOWICZ20191}.
In particular, the IoT-Lite ontology described in \cite{bermudez2016iot} builds a core model containing only the main concepts, along with their relationships, to support the most standard queries for IoT solutions.
Originally proposed by the W3C Semantic Sensor Network Incubator group, the SSN ontology has also been revised by \cite{JANOWICZ20191} in the Sensor, Observation, Sample, and Actuator (SOSA) ontology. 
This proposal aims at a lightweight vocabulary including broader concepts with respect to the SSN ontology, with the idea to provide a core model that can be integrated and aligned to other specifications, e.g. the OGC’s Observations and Measurements (O\&M) or the DOLCE-Ultralite (DUL).

In addition to the effort for the definition of suitable models for the sensor networks and IoT, many researchers also focused on modeling environments, where typically IoT sensors are deployed in \cite{balaji2016brick,rasmussen2021bot,bonino2008dogont,reinisch2010thinkhome}.
One of the first contribution in this direction is represented by the DogOnt ontology \cite{bonino2008dogont}.
DogOnt focuses on a house automation scenario with the objective of modeling home environments, their states and changes, to enabling inter-operation mechanisms and intelligence for more complex actions. This ontology models both architectural elements and controllable ones.
Similarly, the work described in \cite{reinisch2010thinkhome} focuses again on smart homes and proposes the ThinkHome system which comprises a knowledge base and a multi-agent system. One of its main characteristics is the attention to energy efficiency, which distinguishes it with respect to the other proposals.

A step towards our application scenario is made in the proposal of \cite{balaji2016brick}. Here, the authors focus on commercial buildings and propose a common descriptive schema to represent building metadata. Interestingly, the authors consider also sensors and subsystems included in the building, along with their relationships.
More recently, the authors of \cite{rasmussen2021bot} proposed the Building
Topology Ontology (BOT) to support the exchange of information related to building life-cycles among the actors of the Architecture, Engineering, Construction, Owner and Operation industry and according to the Building Information Modelling (BIM) methodology.
The authors show how their ontology, combined with other existing ones (such as, ontology modeling sensors, observations and IoT devices), can support existing applications to make them inter-operable and shareable among interdisciplinary stakeholders. 

In the context of the Industry 5.0, another important aspect is related to the organization of the industry itself, the roles of the human resources, their hierarchies and privileges, and so forth.
When it comes to semantically modeling the concepts involved in an organization, a reference ontology is the Organization Ontology (ORG) \cite{world2014organization}, aimed at modeling organizational structures and related information through the concepts of organizations, their actors, activities and roles.

Through the availability of domain specific ontologies, researchers and industries have started to adopt more and more knowledge graph-based solutions to realize flexible and homogeneously integrated systems \cite{hogan2021knowledge}.
This is true especially for the Internet of Things domain, in which the intrinsic heterogeneity of devices and standards requires the construction of suitable integration solutions to allow for a flexible and fruitful exploitation of the produced data and services \cite{le2016graph,xie2020multilayer,liu2022knowledge}.
For instance, in \cite{le2016graph} the authors propose a system, called Graph of Things (GoT), which uses an RDF data model to build a Knowledge Graph guaranteeing a
unified representation for both stream data and static ones.
The authors of \cite{xie2020multilayer}, instead, adopt the Knowledge Graph technology to build a middleware for supporting applications leveraging IoT data.
Indeed, due to the high heterogeneity of IoT devices and standards, the so called IoT Knowledge Graph enables uniform management of the IoT devices involved. 
Finally, the work described in \cite{liu2022knowledge} focuses on an Cyber-Physical Production System equipped with an Industrial IoT (IIoT). The huge number of interconnected devices (sensors, actuators, and edge computing devices) produce massive heterogeneous multi-dimensional data.
Therefore, seeking an effective data representation strategy, the authors propose a multi-layer knowledge graph, which includes not only data produced by IoT devices, but also production and business processing data. Leveraging the so constructed Knowledge Graph, the authors, then, propose a cognition decision making solution for resource allocation to support manufacturing processes.

Still related to the domain of interest for our proposal, many authors have also focused on the application of Knowledge Graphs to support industrial environments, in which heterogeneous data and multidisciplinary information are typically produced \cite{bader2020knowledge,li2021exploiting}.
The authors of \cite{bader2020knowledge} propose a Knowledge Graph and an underlying ontology to model the standards, norms, specification, requirement and reference frameworks in the context of the Industry 4.0.
Its objective is not to model data produced in an industry following the Industry 4.0 paradigm, but rather to provide a unified data model for all the legislative and normative ecosystems around it.
In \cite{li2021exploiting}, instead, the authors focus on the  exploitation of the Knowledge Graph technology to support the development of products and service innovation in the industry domain.
To do so, they survey the related literature, up to early 2021, to identify promising solutions exploiting Knowledge Graphs in the industrial contexts.
The authors conclude their analysis by showing the current existence of a gap in the fruitful exploitation of this technology and advocate for the development of novel Knowledge-Graph based solutions.

Despite many works have been focused on the adoption of semantic modeling and Knowledge Graphs to support data integration in both the IoT and industry domains, the advent of the Industry 5.0 and the Internet of Everything paradigms introduces new challenges that, to the best of our knowledge, none of the existing solutions consider and address.
The shift towards a more complex scenario, in which humans, as main actors actively producing data through their Body Sensor Networks, collaborate with smart and semi-autonomous devices, production chains, and business processes, requires the re-design and adaptation of existing data models.
This is the objective of our proposal, which builds upon some of the existing solutions described in this section, and extends them to embrace such a new promising industrial ecosystem.

\section{Reference scenario}
\label{sec:scenario}

\begin{figure}
    \centering
    \includegraphics[width=.45\textwidth]{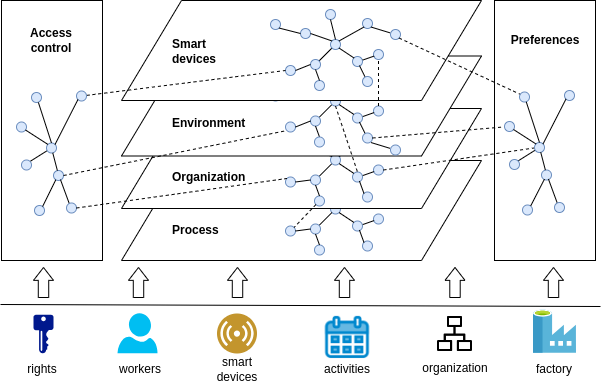}
    \caption{Overview of the information modeled in the framework.}
    \label{fig:overview}
\end{figure}

\noindent In this section, we discuss how information in the framework is categorized based on its typology and the object it represents.
Firstly, we recognize the main perspectives that are intertwined in an IoE environment, as also summarized in Figure \ref{fig:overview}:
\begin{itemize}
    \item the \emph{smart objects} perspective focuses on smart devices composed by one or more systems. In turn, each of them can be composed of a number of other devices such as sensors and actuators. Their technical capabilities, operating ranges, configurations and locations are described;    
    \item the \emph{environment} perspective, which focuses on the enterprise environment and its structure in sub-components, e.g. buildings, floors, rooms;
    \item the \emph{organization} perspective, which includes the organizational structure in terms of sub-organizations and units, and the reporting structure for the employees, where they are located and what roles are defined for them;
    \item the \emph{process} perspective, which details processes in terms of activities in which employees are engaged.
\end{itemize}
Furthermore, two relevant perspectives can be identified that are orthogonal to the mentioned ones, namely:
\begin{itemize}
\item \emph{preferences} of employees related to environmental or smart devices' parameters;
    \item \emph{access control}, which focuses on what rights are associated with each organization role. Rights can be defined of three types with decreasing granularity: on environment, which apply on all systems located in the same place, on smart objects, which apply to all its sub-systems, and on specific systems. 
    
    Furthermore, rights can be transferred among agents (employees or smart objects) in the execution of an activity/process, through different types of relations, namely {\em delegation} or {\em collaboration}. In the former case, the existing rights of the delegated agent for the delegated activity are replaced with those of the delegating agent. Delegation can happen only if the delegated agent is not involved in another activity. On the other hand, in the latter case, the rights already owned by the agent are combined with those of the collaborating one.
\end{itemize}

The data model of the framework is specifically aimed to: (i) provide a model of the IoE (including buildings, agents, roles, devices, smart objects and processes), (ii) provide a characterization of smart objects and devices in terms of their technical specifications and capabilities, (iii) define rules and constraints for access management and (iv) store values produced by smart devices and employees including recorded measurements, environmental parameters, actions and activities performed by devices and employees.
Aspects (i, ii, iii) correspond to the deployment, planning, and configuration of the IoE, while aspect (iv) is concerned with capturing and analyzing the real-time execution and performance of tasks, including the storage of values generated by smart devices and employees.

In the context of Industry 5.0, data management necessitates a foundation built on flexible, interoperable, and easily maintainable structures that adhere to standards in model representation. 
While for aspect (iv) we refer to standard technologies for data storage of IoT devices, this work adopts a Knowledge Graph model for aspects (i, ii, iii) in alignment with established best practices from related literature. On top of it, a set of logical rules are defined in order to represent constraints, e.g. for access control, that cannot be directly represented in the graph model.

Knowledge Graphs are today widely adopted at an enterprise level for their capability of providing a standardized way of representing and organizing data from multiple sources into a unified structure. 
As such, this model enhances data interoperability in scenarios characterized by diverse and heterogeneous information, by referring to shared vocabularies and a structure that can be understood and shared by different systems and applications.
The use of graphs also simplifies data querying and analysis by facilitating the exploration of relationships between entities, hence reducing the time and effort required to find and integrate relevant information.

In particular, we refer to a graph model rooted in RDF (Resource Description Framework) \cite{beckett2004rdf}, a W3C standard for web-based data modeling, storage, and interchange that provides a simple way to describe resources and their relationships, based on the notion of triple. 
Commonly used for (meta)data integration and knowledge representation, RDF is often extended by RDFS (RDF Schema) \cite{brickley2014rdf}, which enhances the vocabulary for describing data structures, including classes, properties, and sub-class/super-class relationships, enabling precise and expressive resource descriptions.
Additionally, OWL (Web Ontology Language) \cite{mcguinness2004owl}, an extension of RDFS, introduces formal semantics for added expressiveness in describing relationships among resources. 
In this work, RDFS and OWL are used for the definition of the SemIoE ontology, which serves as a schema for the platform Knowledge Graph and is introduced in the next section.

\subsection{Example Scenario}
\label{sub:example}
\begin{figure}
    \centering
    \includegraphics[width=.48\textwidth]{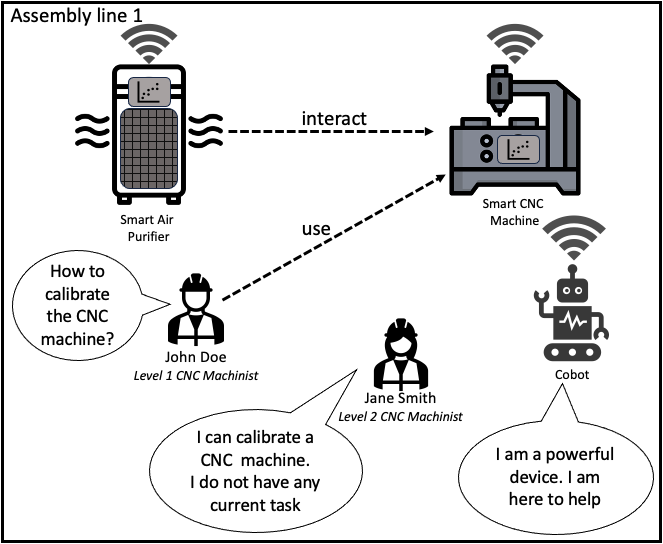}
    \caption{A graphical representation of the considered example scenario.}
    \label{fig:exampleScenario}
\end{figure}

In the following, we describe an example scenario that will be used throughout the remaining of this paper to help the reader understanding our proposed solution.
As sketched in Figure \ref{fig:exampleScenario}, the considered scenario involves an industrial IoE. Here, two employees, namely ``John Doe'' and ``Jane Smith'', are both located at the site ``assembly line 1'', having the role of ``CNC machinist'' with experience {\em level 1} and {\em level 2}, respectively. Both roles involve the right to read from a ``CNC machine'', which is a smart object located in the same site. 
In our example, the ``CNC machine'' is a smart milling machine providing access to its services and data through an underlying IoE network.
Moreover, a ``Cobot'', a collaborative robot, and a smart ``air purifier'' are available on the same site.
Each smart object declares a set of capabilities that it can exploit to carry out tasks.

The considered IoE embraces the principles of autonomy, self-optimization and self-healing, therefore, each involved entity can autonomously interact with other actors to support the activities of the industry they belong to.
In particular, the self-healing mechanism is equipped with a collaborative anomaly detection strategy based on the approaches described in \cite{aramini2022enhanced,arazzi2023fully,arazzi2023novel}.
According to this solution, smart devices of the network monitor each other by learning the expected behavior of their peers (Behavioral Fingerprinting) through lightweight deep-learning models. Once such a typical behavior is known, a peer can verify the presence of anomalies by applying the underlying model to newly generated data.
This is the case of the air purifier constantly interacting with the CNC machine to activate suitable air filtering when the smart milling machine is working on specific materials.
Therefore, it is also in charge of identifying potential anomalies of the CNC machine using the aforementioned Behavioral Fingerprinting solution.

The employee John, instead, is assigned the task of configuring a new contour milling project for the CNC machine.
Such a task would imply a calibration step that requires the support of a more experienced (level 2) employee.
The role of John is a transferable one, so he can engage other users to give him support on his tasks.
Jane, the other employee in this example, is a level 2 machinist and, therefore, has enough experience for advanced configurations of CNC machines. Moreover, her more advanced role allows for advanced tasks, such as calibrating a CNC machine.

\section{Ontological model for IoE}
\label{sec:Ontology}
\noindent This section aims to define the schema of SemIoE, an OWL2 ontology designed to represent agents, smart objects and other entities within an Internet of Everything (IoE) ecosystem. 

The ontology reuses and integrates, according to the best practices in ontology engineering, a number of existing ontological models, as also summarized in Figure \ref{fig:bridge_ontology}.
In the following, we first describe the imported ontologies in Subsection \ref{subsec:external_ontologies}, then in Subsection \ref{subsec:integration} we introduce the set of classes and relations capable to provide the needed connections among the different modules.
We refer to classes and properties from the imported ontologies through the prefixes and corresponding namespaces reported in Table \ref{tab:ontologies}. The namespace of SemIOE is the URI \href{http://w3id.org/semioe\#}{http://w3id.org/semioe\#}, while the 
prefix \emph{ioe} is used for classes and relations. Please note that the URI is a persistent identifier which enables content negotiation. Adhering to best practices for Linked Data, it allows to serve the ontology specification in HTML or the serialization in various formats depending on the accepted MIME type of the HTTP request.

\begin{table}[ht]
\centering
\footnotesize
\begin{tabular}{|l|l|l|}
\hline
\textbf{Ontology}                         & 
\textbf{Prefix} & \textbf{Namespace}                                                 \\
\hline
Semantic Sensor    & ssn                   & http://www.w3.org/ns/ssn/ \\
Network & ssn-system & 
http://www.w3.org/ns/ssn/systems/ \\ \hline
Sensor, Observation, & \multirow{2}{*}{sosa} & \multirow{2}{*}{http://www.w3.org/ns/sosa/} \\
Sample and Actuators                &                       &                                             \\ \hline
Building Topology      & \multirow{2}{*}{bot}  & \multirow{2}{*}{https://w3id.org/bot\#}     \\ 
 Ontology & & \\
\hline
Org ontology                      & org                   & http://www.w3.org/ns/org\#               \\ \hline
\end{tabular}
\vskip\floatsep 
\caption{Prefixes and namespaces for the imported ontologies.}
\label{tab:ontologies}
\end{table}

\subsection{External ontologies}
\label{subsec:external_ontologies}
\noindent Following widely adopted best practices on reusing ontological models, the development starts from the identification of relevant existing vocabularies/ontologies focusing on specific aspects of the IoE, which are detailed in the following.

\subsubsection{Semantic Sensor Network (ssn) and Sensor, Observation, Sample, and Actuator (sosa)}
\label{subsec:ssn_ontology}
the purpose of the SSN ontology \cite{compton2012ssn} is to describe devices in terms of capabilities, measurement processes, observations and deployments.

In SSN, a \emph{ssn:System} is an abstraction for a physical device which can contain other systems.

A system is described in terms of a set of \emph{ssn-system:SystemCapability}, (e.g. accuracy, drift, frequency,  precision, response time), which is a subclass of \emph{ssn:Property} and describes its capabilities in various \emph{ssn-system:Condition}s. 

On the other hand, the SOSA ontology \cite{JANOWICZ20191} provides a lightweight core for SSN, which helps in broadening the target audience and application fields, e.g. by considering both \emph{sosa:Sensor} and \emph{sosa:Actuator} as subclasses of system in a coherent framework with a flexible representation.

\subsubsection{Building Topology Ontology (bot)}
\label{subsec:bot}
developed by the W3C Linked Building Data Community Group, the Building Topology Ontology \cite{rasmussen2021bot} is a minimal OWL DL ontology,  aimed at defining relationships between the sub-components of a generic building.

The main class is the \textit{bot:Zone} which is defined as a part of the physical/virtual world that is inherently both located in this world and has a 3D spatial extent. Subclasses are of different types: a \textit{bot:Site} is an area containing one or more \textit{bot:Building}s, i.e. an independent unit of the built environment with a characteristic spatial structure. This last includes one or more \textit{bot:Storey}, namely a level part of a building, which in turn contains \textit{bot:Space}s, i.e. limited three-dimensional extents defined physically or notionally.
Besides containment, other spatial relations are defined, such as adjacency and intersection. Several classes are aligned to upper level ontologies such as DOLCE-UltraLite ontology (DUL) and domain ontologies, among which BRICK, DogOnt, ThinkHome.

\subsubsection{Organization ontology (org)}
\label{subsec:org_ontology}
published as a W3C Recommendation, the Organization ontology \cite{world2014organization} is aimed to represent organizational structures. It is designed to allow domain-specific extensions to add classification of organizations and roles, as well as extensions to support neighboring information such as organizational activities.
The ontology includes classes and properties to support the representation of information typically reported in organizational charts: (1) the organizational structure, including the decomposition of an \emph{org:Organization} into sub-organizations and \emph{org:OrganizationalUnits}, (2) the reporting structure with \emph{org:Membership}, \emph{org:Role}, \emph{org:Post} and relations among people, (3) location information including \emph{org:Site}, i.e. buildings where the organization (or a sub-part thereof) is located, and (4) the organizational history (e.g. renaming of structures).
Apart from these core concepts, the ontology can be extended to include more detailed information on organizational control structures and flows of accountability and empowerment.

\subsection{Integrated model}
\label{subsec:integration}

\begin{figure*}[ht]
    \centering
    \includegraphics[width=.7\linewidth]{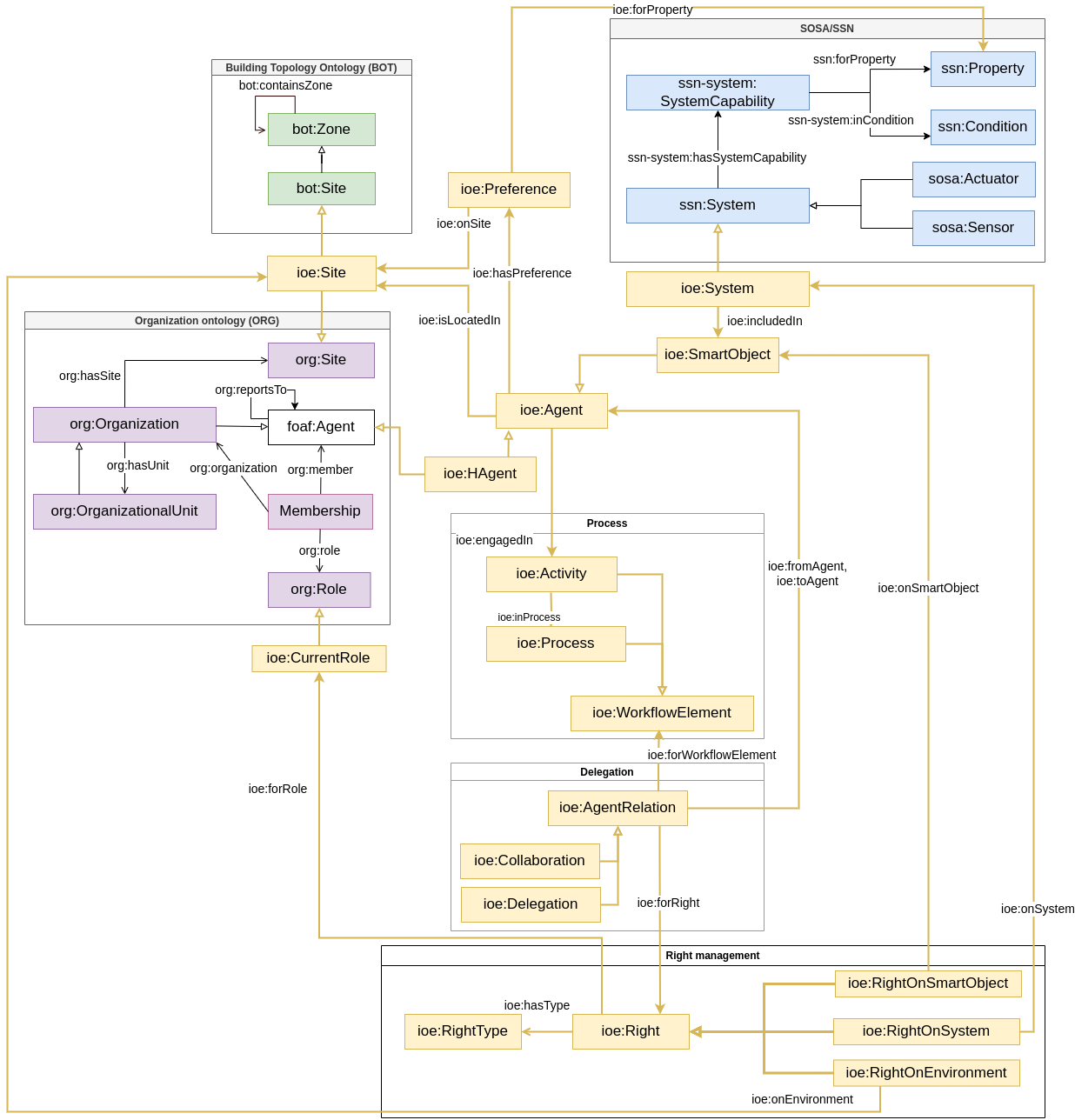}
    \caption{Integrated schema of the SemIoE ontology.}
    \label{fig:bridge_ontology}
\end{figure*}

\noindent The ontological modules described above have been integrated through a bridge ontology which is graphically represented in Figure \ref{fig:bridge_ontology}. On the one hand, it allows to define the needed connections and alignments between relevant classes of different ontologies for the realization of a integrated model of the IoE environment.
On the other hand, the bridge ontology defines further classes and properties to represent concepts that are relevant for the IoE environment (see Section \ref{sec:scenario}) and are not available in external ontologies. As such, the resulting ontology is equivalent to ${\cal O}_{SemIoE}$ $\equiv {\cal O}_{Bridge} \sqcup {\cal M}_{sosa/ssn} \sqcup {\cal M}_{bot} \sqcup {\cal M}_{org}$, where $\cal M$ stands for the ontological mappings to the corresponding ontology.

The main class in the bridge ontology is \emph{ioe:Agent}, which represents an agent in the IoE as a generalization of both a \emph{ioe:HAgent}, i.e. a human agent, and a \emph{ioe:SmartObject}, i.e. a smart device. Such a distinction allows to associate properties to each subclass separately. Both classes are intended to be extended with further properties for future usage. 

The class \emph{ioe:SmartObject} includes at least a \emph{ioe:System}, which represents a device. According to our modeling approach, a smart object encompasses both a simple device, e.g., a sensor, and a more complex smart device, i.e., a smart watch. While in the former case the smart object would include a single system, in the second case it can include multiple systems (property \emph{ioe:includedIn}). While technical capabilities of a system can be expressed through the class \emph{ssn-system:SystemCapability}, higher level functionalities provided by the system can be represented by extending this concept.

An agent is located at an \emph{ioe:Site} of the organization. Furthermore, an agent can be engaged in an \emph{ioe:Activity}. This holds for both a human agent and a smart object, which in some cases may possess enough capabilities to perform activities in the organization. In the execution of an activity, multiple agents can be engaged, e.g. if a human agent is performing an activity using a CNC machine, both the former and the latter are engaged in the same activity. 

The representation of the process perspective is realized by representing activities through the class \emph{ioe:Activity}, which is part of a \emph{ioe:Process}, and both are subclasses of \emph{ioe:WorkflowElement}\footnote{More expressive ontological representations of processes and workflows have been proposed in the literature (see \cite{diamantini2023process} for a discussion).}.

The access control perspective includes the representation of rights through the class \emph{ioe:Right}, which defines a right of a particular \emph{ioe:RightType}, e.g. read, write, update, delete, and is associated with a \emph{ioe:CurrentRole}. It includes as subclasses:
\begin{itemize}
\item \emph{ioe:RightOnSystem} is connected to the specific \emph{ioe:System} which is the target of the right through property \emph{ioe:onSystem};
\item \emph{ioe:RightOnSmartObject}, on the other hand, specifies a right which applies to a whole smart object (through property \emph{ioe:onObject}), including all the (sub)systems included thereby;
\item \emph{ioe:RightOnEnvironment} defines, through property \emph{ioe:onEnvironment}, a right on all smart objects which are \emph{ioe:locatedIn} a given \emph{ioe:Site}.
\end{itemize}

Finally, the representation of preferences is realized through the class \emph{ioe:Preference}, which can be expressed by an \emph{ioe:Agent}, either a human agent and a smart object. A preference is defined on a certain \emph{ssn:Property} measured in a given \emph{ioe:Site} and specifies (through the data property \emph{ioe:hasPreferenceValue}) the preferred value for the property.
For what concerns the relations among agents, the class \emph{ioe:AgentRelation} defines a relation of \emph{ioe:Collaboration} or \emph{ioe:Delegation} between two \emph{ioe:Agent}s (relations \emph{ioe:fromAgent}, \emph{ioe:toAgent}). The relation needs to specify the activity/process it is intended for, through property \emph{ioe:forWorkflowElement}. As a result of the establishment of the relation, one or more new \emph{ioe:Right}s are associated with the new \emph{ioe:CurrentRole} of the target of the collaboration/delegation. The role specifies its temporal validity through starting and ending time (data properties \emph{ioe:startTime} and \emph{ioe:endTime}) and whether it can be transferred to other agents (data property \emph{ioe:isTransferable}).

The ontology includes the following alignments among classes of the bridge ontologies and classes of external modules:
\begin{itemize}
\item the \emph{ioe:System} class is aimed to represent a system as a subclass of a \emph{ssn:System} in the SSN ontology (\emph{ioe:System} $\sqsubseteq$ \emph{ssn:System}). As such, the properties of the class are inherited. In particular, an \emph{ioe:System} can be defined compositionally, through the property \emph{ssn:hasSubSystem}.

\item the class \emph{ioe:HAgent} is a subclass of \emph{foaf:Agent} (\emph{ioe:HAgent} $\sqsubseteq$ \emph{foaf:Agent}). As a consequence, the properties of the class are inherited. In particular, a human agent has an \emph{org:Role};
\item the class \emph{ioe:CurrentRole} is a specialization of \emph{org:Role} (\emph{ioe:CurrentRole} $\sqsubseteq$ \emph{org:Role});

\item the class \emph{ioe:Site} specializes both \emph{bot:Site} and \emph{org:Site} (\emph{ioe:Site} $\sqsubseteq$ \emph{org:Site}, \emph{ioe:Site} $\sqsubseteq$ \emph{bot:Site}). In this way, the properties of such classes are inherited by \emph{ioe:Site}, e.g. to enable the containment relation among sites that is defined in the BOT ontology (\emph{bot:containsZone}). The alignment also allows to recognize that an organizational unit, e.g. an office, is the same \emph{ioe:Site} where an agent is located.

\end{itemize}

The ontology has been implemented through Protégé (version 5.6.3)\footnote{https://protege.stanford.edu/}. It includes 136 axioms, 20 classes and 10 object properties. Its consistency has been checked with the reasoner HermiT (version 1.4.3.456)\footnote{http://hermit-reasoner.org/}. The serializations of the ontology in the Turtle \cite{beckett2004rdf} and RDF/XML formats are available at the URL \href{http://w3id.org/semioe}{http://w3id.org/semioe}. 

\subsection{Example}
In this example we show the serialization in Turtle of a simple case scenario using classes and relations from the SemIoE ontology. The scenario is the one described in Section \ref{sub:example} and, in this case, John asks for the collaboration of Jane in the execution of an activity ``calibration'' he is performing on the CNC machine.
The resulting Turtle serialization is reported in Listing \ref{ls:turtle}.

\begin{lstlisting}[morekeywords={SELECT,WHERE},
   basicstyle=\ttfamily\scriptsize, frame=single, label={ls:turtle}, caption={A Turtle serialization using the SemIoE ontology.}, captionpos=b]
:assembly_line_1 rdf:type ioe:Site .

:calibration rdf:type ioe:Activity .

:jane_smith rdf:type ioe:HAgent ;
            ioe:locatedIn :assembly_line_1 .

:john_doe rdf:type ioe:HAgent ;
          ioe:engagedIn :calibration ;
          ioe:locatedIn :assembly_line_1 .

:CNC_machine rdf:type ioe:SmartObject ;
             ioe:locatedIn :assembly_line_1 .

:CNC_machinist_level_1 rdf:type ioe:CurrentRole ;
                  ioe:isTransferable "true"^^xsd:boolean .

:CNC_machinist_level_2 rdf:type ioe:CurrentRole .

:membership_1 rdf:type org:Membership ;
              org:member :john_doe ;
              org:role :CNC_machinist_level_1 .

:membership_2 rdf:type org:Membership ;
              org:member :jane_smith ;
              org:role :CNC_machinist_level_2 .
              
:agent_relation_1 rdf:type ioe:Collaboration ;
                  ioe:forWorkflowElement :calibration ;
                  ioe:fromAgent :john_doe ;
                  ioe:toAgent :jane_smith .

:configure rdf:type ioe:RightType .

:read rdf:type ioe:RightType .

:right_read_CNC_machine rdf:type ioe:RightOnSmartObject ;
                   ioe:forRole :CNC_machinist_level_1 ,
                               :CNC_machinist_level_2 ;
                   ioe:hasType :read ;
                   ioe:onSmartObject :CNC_machine .

:right_config_CNC_machine rdf:type ioe:RightOnSmartObject ;
                   ioe:forRole :CNC_machinist_level_1 ;
                   ioe:hasType :configure ;
                   ioe:onSmartObject :CNC_machine .

\end{lstlisting}

\section{Services for the IoE}
\label{sec:services}

\noindent In this section, we define different types of service providing support to the execution of specific activities within a Industry 5.0 scenario. Services are built on top of the semantic layer of the platform, which includes the SemIoE ontology and the Knowledge Graph.

Hereby, we focus on the formulation of advanced support services assuring the fulfillment of activities possibly involving, at least in some cases, multiple agents that cooperate for their execution based on their roles.
The following services will be detailed below: {\em (i)} IoE Access Control, {\em (ii)} IoE Collaboration, {\em (iii)} IoE Secure Delegation, and {\em (iv)} IoE Environment Setting.

\subsection{IoE Access Control}
\noindent In our scenario, we assume that the IoE facilities of the industry are restricted to agents with a specific role.
Therefore, an IoE Access Control service is necessary to prevent unauthorized users from accessing the data collected by the sensors.

The SemIoE ontology models each environment of the factory as an \emph{ioe:Site} composed by a set of sub-sites, e.g., a building or a room.
In each site, processes are regulated by a series of \emph{ioe:System}s (e.g., sensors) included in \emph{ioe:SmartObject}s.
A specific \emph{ioe:HAgent},  when entering the site, will be allowed to have access to only the subset of \emph{ioe:System}s that are compatible with his/her \emph{org:Role}.

This service is crucial for a smart factory because the recent scientific literature reports different types of attack that can leverage the manipulations of smart sensors to orchestrate distributed attacks on the entire \emph{ioe:SmartObject}s \cite{hoffman2009survey, vishwakarma2020survey}.

As reported in Algorithm \ref{alg:accessControl}, when an agent $a$ enters an environment $e$ to perform an activity on a system $s$ with a specific right type $t$ (e.g., read), the framework retrieves the information that characterizes the $role$ of the agent and the required rights to access the system $s$.

In particular, the framework checks if the agent $a$ and the specific system $s$ are in the same environment $e$ (line 1).
If confirmed, the framework retrieves the rights $RH=\{rh_1,\ldots,rh_n\}$ corresponding to the role $r$ of the agent $a$ (line 3-4). It cycles through them (line 5) and returns $True$ if the agent has the rights on the system $s$ of type $t$ (lines 6-8), or if the agent has a right on environment $e$ of type $t$ (lines 11-13). Otherwise,  the framework returns $False$ if these conditions are not verified.

\paragraph*{\bf Use case example} let us consider again the scenario of Section \ref{sub:example}, when the agent John enters the site and accesses the framework to configure the new milling task on the CNC machine. The agent will be identified with a specific role characterized by a set of rights, and the rights associated with the role will then be verified for compliance with the requested task.
With a positive outcome, the framework allows the agent to get control of the CNC machine and provides the permissions to access the precision cameras, proximity and pressure sensors to configure it for the new milling task.

\begin{algorithm}[ht] 
\footnotesize
\caption{Access Control \label{alg:accessControl}}
\begin{algorithmic}[1]
\Require{a (Agent), s (System), e (Site), t (RightType)} 
\If {locatedIn(s)!=locatedIn(a)} 
\Return False
\Else {
\State $r$: Role = getRole(a)
\State $RH$: list$<$Right$>$ = getRights(role)
\ForEach{$rh_i \in RH$}
\If {RightOnSystem($rh_i$)}  
\State S = getSystemsFromRight($rh_i$)
\If {$s \in S \wedge t$==getType($rh_i$)}
 \Return True
\EndIf
\EndIf\
\If {RightOnEnvironment($rh_i$)} 
\State $env$ = getEnvironmentFromRight($rh_i$)
 \If {$env$ == $e \wedge t$==getType($rh_i$)}
 \Return True
\EndIf
\EndIf

\Return False
\EndFor
}
\EndIf

\end{algorithmic}
\end{algorithm}

\subsection{IoE Collaboration}

\noindent The second service enables the collaboration between multiple agents (\emph{ioe:HAgent}) with roles (\emph{org:Role}) of different level. In particular, the service considers the case in which two agents, say $a_1$ and $a_2$, with different roles, need to collaborate to complete an activity.
The requirement is that the two agents must be co-located in the same {\em ioe:Site}. 
The service ensures that an agent can require the collaboration of another agent and can grant access to the same facilities in the IoE.
In practice when $a_1$ has to complete an activity in a given time $t$, it can leverage the cooperation of $a_2$ by temporarily extending its privileges ({\em ioe:Right}) to grant access to the facilities of the IoE accessible by $a_1$.

In particular, as described in the Algorithm \ref{alg:collaboration}, when the agent $a_1$ asks to collaborate with another agent with a different role, say $a_2$, the framework has to check if the role $r_1$ of agent $a_1$ is transferable.
To do so, the framework uses the function $isTransferable$ (line 3). If the function returns $True$ the procedure continues, otherwise it returns $False$ and the process ends. 
If the outcome is positive the collaboration procedure starts.
In practice, the framework retrieves from the roles $r_1$ and $r_2$ the list of rights $RH_{r_1}$ and $RH_{r_2}$ associated with them (lines 4-5). 
Then, the framework can create a new temporary role characterized by a list of rights obtained by the union between $RH_{r_1}$ and $RH_{r_2}$ and a period in which this new temporary role is valid (lines 6-7). 
The new temporary role is now ready to be assigned to the agent $a_2$ (line 8).

\paragraph*{\bf Use case example (continued)} now John has to proceed with the calibration of the CNC machine, which is a required step before being able to execute the milling task. This task requires the collaboration of a level 1 employee to avoid possible errors.
Through the framework, he can request the collaboration of another agent, the human employee Jane, in this case.
Then, the framework verifies if the rights of John are transferable.
With a positive response, the framework can now generate a new temporary role for Jane with the required rights to fulfill the task assigned to John.

\begin{algorithm}[ht] 
\footnotesize
\caption{IoE Collaboration \label{alg:collaboration}}
\begin{algorithmic}[1]
\Require{$a_1$ (Agent), $a_2$ (Agent), $st$ (Start Time), $et$ (End Time)}  
\State $r_1$: Role = getRole($a_1$)
\State $r_2$: Role = getRole($a_2$)
\If{isTransferable($r_1$)}
 \State $RH_{r_1}$: list$<$Right$>$ = getRights($r_1$)
 \State $RH_{r_2}$: list$<$Right$>$ = getRights($r_2$)
 \State $RH_c$ = $RH_{r_1} \bigcup RH_{r_2}$
\State $r_c$ = createTempRole($RH_c$, $st$, $et$) 
\State $a_2$.$r_2$ = assignRole($r_c$, startTime=$st$, endTime=$et$)
\EndIf
\end{algorithmic}
\end{algorithm}

\subsection{IoE Secure Delegation}
   
\noindent Analogously to the previous service, a given \emph{ioe:Agent}, say $a_1$, can leverage the availability of another \emph{ioe:Agent}, say $a_1$, to complete a critical activity on $a_1$'s behalf.

The idea behind this service is to allow the agents to share access to resources, devices or data, available to them, with an adequate security level. 

In practice, following Algorithm \ref{alg:IoEDelegation}, due to the impossibility of executing an activity, agent $a_1$ can delegate the process to a second agent $a_2$ to carry out a given activity $ac$, in time.
For this reason, the delegation $d$ must be elapsed in a given period of time defined by a start time $st$ and an end time $et$.
In particular, the current roles $r_1$ and $r_2$ of the agents are retrieved using the function $getRole$ (lines 2-3).
Before proceeding with the process of delegation, the framework checks if the role $r_1$ is transferable to other agents (line 4).

If the outcome is positive, is now possible to retrieve the smart objects $O_{ac}$ involved in the execution of activity $ac$ together with $a_1$, using $getObjectsEngagedInActivity$ (line 7).
Then, the rights associated with these objects can be derived with the function $getRightsOnObjects$ (line 8).
After that, the list of objects can be used to get the involved systems $S_{ac}$ using $getSystemsIncludedIn$ and all the associated rights with $getRightsOnSystems$ (lines 9-11).
At this point, the algorithm identifies the subset of rights $RH_d$, intersection between the rights $RH_{r_1}$ associated with the role $r_1$ of the agent $a_1$ and the entire set of rights, namely $RH_{OS_{ac}}$ (line 12), on the objects $O_{ac}$ and systems $S_{ac}$ involved.

The retrieved set of rights $RH_d$ can now be used to generate the delegation rights (line 13). 
The obtained temporary rights are used to generate the new delegation role $r_d$ (line 14), which is then assigned to $a_2$ with a limited life span limited by $st$ and $et$ (line 15).

\paragraph*{\bf Use case example (continued)} the reliability of the CNC machine is guaranteed by the constant anomaly detection performed by the air purifier in the scenario of \ref{sub:example}.
However, this device may not be able to initialize and execute a behavioral model, as this task requires the training and inference of a deep learning model \cite{aramini2022enhanced,arazzi2023fully, arazzi2023novel}.
For this reason, it can ask through the framework if a more capable device is available for managing a behavioral fingerprinting model using its data. 
The framework identifies the cobot as a device showing high computational capability.
Similarly to the collaboration service, the delegation service proceeds by generating a new role to assign to the identified powerful device that now will act on behalf of the air purifier to complete the target task.

\begin{algorithm}[ht] 
\footnotesize
\caption{IoE Secure Delegation \label{alg:IoEDelegation}}
\begin{algorithmic}[1]
\Require{ $a_1$ (Agent), $a_2$ (Agent), $ac$ (Activity), $st$ (Start Time), $et$ (End Time)} 
\State $S_{ac}$:list$<$System$>$ = []
\State $r_1$: Role = getRole($a_1$)
\State $r_2$: Role = getRole($a_2$)
\If{isTransferable($r_1$)}

\State $RH_{r_2}$: list$<$Right$>$ = getRights($r_2$)
\State $RH_{r_1}$: list$<$Right$>$ = getRights($r_1$)

\State $O_{ac}$:list$<$Object$>$ = getObjectsEngagedInActivity($ac$)
\State $RH_{O_{ac}}$: list$<$Right$>$  = getRightsOnObjects($O_{ac}$)
\ForEach{$o_i \in O_{ac}$}
\State $S_{ac} \leftarrow $ getSystemsIncludedIn($o_i$)
\EndFor
\State $RH_{OS_{ac}}$: list$<$Right$>$  = getRightsOnSystems($S_{ac}$) $\bigcup RH_{O_{ac}}$

\State $RH_d$: list$<$Right$>$ =$RH_{OS_{ac}} \bigcap RH_{r_1}$

\State $r_d$ = createTempRole($RH_d$, $st$, $et$) 
\State $a_2$.$r_2$ = assignRole($r_d$, startTime=$st$, endTime=$et$)
\EndIf
\end{algorithmic}
\end{algorithm}

\subsection{IoE Environment Setting}
    
\noindent This service focuses on the adaptation of the environmental parameters of a \emph{ioe:Site}, e.g., a room, according to the preferences specified by an \emph{ioe:Agent}. 

The service can access the information from the multiple \emph{sosa:Sensor}s inside the site, and can leverage the actuators ({\em sosa:Actuator}) to adapt the environmental parameters of the room on the basis of the preferences of the \emph{ioe:HAgent} that enters the site. 
This service is aimed at improving the productivity of an agent adjusting its working environment, such as, for example, the room temperature, the lightning level, and so forth, matching its preferences automatically.
As presented in Algorithm \ref{alg:IoEEnvSet}, when an agent $a_1$ accesses a site $e$ the framework retrieves the system's properties preferences $Pref_{a1}$ specified by agent $a_1$ (line 2).
Then, the framework retrieves the systems $S_{e}$ and their associated properties $P_{S_e}$ in the site $e$ (lines 3-6) and applies to them the preferences $Pref_{a1}$ of the agent $a_1$ (line 9).

\paragraph*{\bf Use case example (completed)} 
After the calibration, the CNC Machine is ready to carry out the new milling task. At this point, when activated, it can alter the environment setting to require the increase of the power of the filtration system of the air purifier to absorb the produced particles and preserve the air quality of the room.

\begin{algorithm}[ht] 
\footnotesize
\caption{IoE Environment Setting \label{alg:IoEEnvSet}}
\begin{algorithmic}[1]
\Require{ $a_1$ (Agent), $e$ (Site)}
\State $P_e$ : list$<$Property$>$ = []
\State $Pref_{a1}$ : list$<$Preference$>$ = getEnvPreferences($a_1$)
\State $O_{e}$ : list$<$SmartObject$>$ = getObjectsLocatedIn($e$)

\ForEach{$o_i \in O_e$}
\State $S_i$ : list$<$System$>$ = getIncludedSystems($o_i$)
\State $P_e \leftarrow$  getSystemsProperties($S_i$)
\EndFor

\ForEach{$p_i \in P_e$}
\State $p_i$ = applyPreferences($p_i$, $Pref_{a1}$)
\EndFor

\end{algorithmic}
\end{algorithm}

\section{Implementation and Discussion}
\label{sec:eval}
\noindent 

In this work, we refer to a basic IoT infrastructure comprising a collection of smart objects, with some permanently deployed in the environment and others worn by employees. These devices possess the capability to monitor various parameters and execute actions. Generated data can be optionally stored in a Cloud infrastructure, typically through standard protocols (e.g., MTTQ, HTTP) or directly retrieved from the device, depending on their peculiarities and capabilities. For some type of information, e.g. on personal health-related aspects, data is stored and consumed locally on the devices and cannot be accessed from outside.
The Knowledge Graph of the IoE is defined on top of the IoT infrastructure, enabling the integration of multiple, heterogeneous information by relying on classes and relations defined in the SemIoE ontology.

To support data storage and retrieval of information from the Knowledge Graph, a catalog of services provides functionalities at different levels, ranging from functions to perform low-granularity operations to advanced capabilities. 
\paragraph{Storage layer} it includes low-level CRUD functionalities operating on specific triples, namely the capability to add (delete) a given triple to (from) the graph. Such functions are provided by a triplestore, i.e. a DBMS specifically tailored to the storage and management of triples. Additionally, generic SPARQL  \cite{sparql} queries can be run on the graph, to support medium- and higher-level functions.

\paragraph{Entity layer} it includes medium-level functionalities to manage graph entities such as agents, smart devices, environments, roles, rights or privileges. These functions typically operate on groups of triples and are categorized in two classes. The first is formed by functions aimed to extract all triples having the given instance as a subject. The following SPARQL query extracts all information centered on a given instance $ins$ of class $entity$, in terms of properties and corresponding objects:

\begin{lstlisting}[captionpos=b, caption=SPARQL query for information on instance \emph{ins} of type \emph{entity}., morekeywords={SELECT,WHERE},
   basicstyle=\ttfamily\small,frame=single]
  SELECT ?p ?o
  WHERE {<ins> a <entity>;
               ?p ?o}
\end{lstlisting}

A second group of functions are devoted to extract specific pieces of information from the graph, to support more advanced functionalities. The following ones are respectively aimed to extract the \emph{ioe:Site} in which an \emph{ioe:Agent} is located (locatedIn), its \emph{ioe:CurrentRole}  (getRole), the rights associated to such a role (getRights) and the systems on which a given \emph{ioe:RightOnSystem} is defined (getSystemsFromRight).

\begin{lstlisting}[captionpos=b, caption=SPARQL query for function \emph{locatedIn()}., morekeywords={SELECT,WHERE},
   basicstyle=\ttfamily\small, frame=single]
  SELECT ?site
  WHERE {<agent> a ioe:Agent;
                 ioe:locatedIn ?site}
\end{lstlisting}

\begin{lstlisting}[captionpos=b, caption=SPARQL query for function \emph{getRole()}., morekeywords={SELECT,WHERE},
   basicstyle=\ttfamily\small,frame=single]
  SELECT ?role
  WHERE {?m a org:Membership;
            org:member <agent>;
            org:role ?role}
\end{lstlisting}

\begin{lstlisting}[captionpos=b, caption=SPARQL query for function \emph{getRights()}., morekeywords={SELECT,WHERE},
   basicstyle=\ttfamily\small,frame=single]
  SELECT ?right
  WHERE {?right a ioe:Right;
                ioe:forRole <role>}
\end{lstlisting} 

\begin{lstlisting}[captionpos=b, caption=SPARQL query for function \emph{getSystemsFromRight()}., morekeywords={SELECT,WHERE},
   basicstyle=\ttfamily\small,frame=single]
  SELECT ?sys  
  WHERE {<right> a ioe:RightOnSystem;
                   ioe:onSystem ?sys} 
\end{lstlisting} 

Further entity level functions can support specific purposes that are useful in the IoE context, e.g. retrieval of resources having a given status. For instance, the following query searches for all human agents in the organization that are available, i.e. that are not currently assigned to some activities.
\begin{lstlisting}[captionpos=b, caption=SPARQL query to retrieve available agents., morekeywords={SELECT,WHERE},
   basicstyle=\ttfamily\small,frame=single]
  SELECT ?agent 
  WHERE {?agent a ioe:HAgent. 
         MINUS {?agent ioe:engagedIn ?act. 
                ?atc a ioe:Activity.}}
\end{lstlisting}

\paragraph{Support layer} it includes high-level functions to implement the advanced management services for the IoE, as described in Section \ref{sec:services}.  

Each group relies on the functionalities provided by the lower level. To give an example, the advanced service \emph{IoE Access Control} relies on the entity-level services \emph{getRole} and \emph{getRights} respectively to extract the role for a given agent and the rights for a given role. Likewise, the service \emph{getRole}, as shown, requires the capability to query triples related to the specified role.
The functionalities provided by each layer are implemented as APIs in the framework, on which the logic of higher-level functions are constructed.

\section{Conclusion}
\label{sec:conclusion}

In this paper, we have focused on a novel reference scenario, in which the Internet of Everything (IoE) technology is adopted as a main driver to foster the transition towards the Industry 5.0 paradigm.
Our investigation started from the analysis of the complexity of the considered scenario, for which the integration of the data produced by the entities of such an industrial IoE appears critical, due to their intrinsic heterogeneity, the magnitude of the surrounding environment, and the complexity of required services and tasks.
For this reason, we developed a semantic model to formalize the entities and the dynamics involved in an IoE-based industry. 
Following the reuse and abstraction design principles, we, hence, proposed a novel ontology, called SemIoE, which is built by extending and complementing consolidated state-of-the-art models, such as the Semantic Sensor Network, the Building Topology Ontology, and the Organization Ontology.
Additionally, to support the activities carried out in our reference Industry 5.0 scenario, we defined and developed a set of IoE services built on top of a Knowledge Graph leveraging SemIoE ontology.
The proposal described in this paper represents a first contribution towards the establishment of new-generation industries embracing the Industry 5.0 paradigm and leveraging the Internet of Everything technology.
In the future, we plan to extend our solution by considering a more refined concept of service provisioning in the IoE, according to which each involved entity can act as both a user and a provider of a set of services.
Moreover, the data produced by specific services may be subject to different security and privacy constraints, thus imposing the necessity to model classes of security and privacy requirements that could extend our ontology.
Finally, we plan to integrate our solution with a human interaction module that will engage with the platform services and the Knowledge Graph, contributing to the creation of a user-centric immersive working environment.

\bibliographystyle{plain}

\end{document}